\documentclass[12pt]{article}
\usepackage{a4wide}

\usepackage[dvipdfmx,hiresbb]{graphicx} 
\usepackage{mediabb}                     

\usepackage{color}

\newcommand{\addedSecond}[1]{{\color{black}#1}}

\newcommand{\al}[1]{\begin{align}#1\end{align}}
\newcommand{\als}[1]{\begin{align*}#1\end{align*}}
\newcommand{\ab}[1]{\left|#1\right|}
\newcommand{\paren}[1]{\left(#1\right)}
\newcommand{\fn}[1]{\!\left(#1\right)}
\newcommand{\sqbr}[1]{\left[#1\right]}

\newcommand{\n}{\text{\sf n}}

\usepackage{amsmath,amssymb}
\newcommand{\df}{\text{d}}

\usepackage{epsf}

\newcommand{\GeV}{\ensuremath{\,\text{GeV} }}

\newcommand{\nn}{\nonumber\\}

\begin{document}

\title{\vbox{
\baselineskip 14pt
\hfill \hbox{\normalsize KUNS-2567}
} \vskip 1cm
\bf \Large Saddle point inflation in string-inspired theory \vskip 0.5cm
}
\author{
Yuta~Hamada,\thanks{E-mail: \tt hamada@gauge.scphys.kyoto-u.ac.jp}~
Hikaru~Kawai,\thanks{E-mail: \tt hkawai@gauge.scphys.kyoto-u.ac.jp}~ 
and Kiyoharu~Kawana\thanks{E-mail: \tt kiyokawa@gauge.scphys.kyoto-u.ac.jp}\bigskip\\
{\it \normalsize
Department of Physics, Kyoto University, Kyoto 606-8502, Japan}\smallskip
}
\date{\today}


\maketitle

\abstract{\noindent \normalsize
The observed value of the Higgs mass indicates the possibility that there is no supersymmetry below the Planck scale and that the Higgs can play the role of the inflaton.
We examine the general structure of the saddle point inflation in string-inspired theory without supersymmetry.
We point out that the string scale is fixed to be around the GUT scale~$\sim10^{16}\GeV$ in order to realize successful inflation. 
We find that the inflaton can be naturally identified with the Higgs field.
}

\newpage

\normalsize

\section{Introduction}
The recently observed particle at ATLAS~\cite{:2012gk} and CMS~\cite{:2012gu} experiments at the Large Hadron Collider (LHC) is consistent with the Standard Model (SM) Higgs with the mass around $125\GeV$.
Up to now, there has been observed no significant deviation from the SM nor a hint of new physics. 
Once the Higgs mass is determined, we have fixed all the parameters in the SM and can extrapolate it up to its ultraviolet (UV) cutoff scale.
In particular, the quadratically divergent bare Higgs mass is found to be suppressed when the UV cutoff is at around the Planck scale~\cite{Bare mass}, see also Ref.~\cite{Jones}.
Furthermore, the quartic Higgs coupling becomes tiny at the same time, see e.g  Refs.~\cite{Bare mass,SM stability}. 
This opens up the possibilities of identifying the Higgs field as the inflaton~\cite{Higgs inflation,Hamada:2015ria}, and of the absence of supersymmetry below the Planck scale. 
%
%
Although non-supersymmetric vacua are ubiquitous in string theory~\cite{nonSUSYpheno,nonSUSY}, their phenomenology has not been well studied.
It becomes important to explore the phenomenology starting from non-supersymmetric theory. 

In this letter, we consider the saddle point inflation scenario starting in string-inspired theory without supersymmetry.
The potential is generated perturtabatively in contrast to the supersymmetric case where the potential comes only non-perturbatively.
Then, we calculate the cosmological parameters by assuming that the potential is tuned in such a way that the first $\n$ derivatives vanish at some point.
The predicted cosmological parameters are consistent with the recent Planck 2015 result~\cite{Ade:2015lrj}.
Furthermore, we can estimate the order of the string scale from the height of the potential that is roughly given by the string scale to the fourth multiplied by the rather small ten-dimensional one-loop factor.

To realize the saddle point, some amount of the fine-tuning is needed. 
This fine-tuning would be achieved by some principles which are beyond the ordinary local field theory, e.g. the multiple point criticality principle~\cite{MPP} and the maximum entropy principle~\cite{MEP}.

This letter is organized as follows.
In the next section, we consider the potential that has a saddle point where the first $\n$ derivatives vanish.
Then we calculate the cosmological parameters of the model.
%
%
In Sec.~\ref{Sec:string}, we estimate the stirng scale 
in the case of the non-supersymmetric heterotic-like string model.
%
In Sec.~\ref{Sec:summary}, we summarize our result.

\section{Saddle point inflation and observables}\label{Sec:saddle point inflation}
We start with a general potential $V$ as a function of an inflaton field $\varphi$.
We will discuss the possibility of identifying it as the SM Higgs in the next section. 
We expand the potential around the saddle point $\varphi_c$ as $\varphi=\varphi_c+\delta\varphi$:
\al{
V
	&=	\sum_{n=0}^\infty{V_c^{(n)}\over n!}{\delta\varphi}^n
	=	V_c+V_c'\,\delta\varphi+{V_c''\over2}\,{\delta\varphi}^2+{V_c'''\over3!}\,{\delta\varphi}^3+\cdots.
		\label{potential expanded}
}
We assume that the first $\n$ ($\geq2$) derivatives vanish at $\varphi_c$:

\al{
V_c'=V_c''=\dots=V_c^{(\n)}=0.
}
Here, we also assume $V_c^{(\n+1)}>0$ ($<0$) for even (odd) $\n$ so that $\varphi$ rolls down from $\varphi_c$ towards $0$.
This is because we are going to identify $\varphi$ as the Higgs field.

%

The slow roll parameters around the saddle point are obtained as
\footnote{
{
If $\varphi_c$ is the only mass scale of theory as the model we will consider in the next section,
we have $V_c^{(n+2)}/V_c^{(n+1)}\sim \varphi_c^{-1}$.
}
Therefore, the condition for the validity of neglecting the higher order terms is
\als{
{V_c^{(n+2)}\delta\varphi^{n+2}\over(n+2)!}
\bigg/
{V_c^{(n+1)}\delta\varphi^{n+1}\over(n+1)!}
       \sim {\delta\varphi\over \varphi_c}
	\ll	1.
}
}
\al{
\epsilon
	&:=	{M_P^2\over 2}\paren{V'\over V}^2
	=	{M_P^2\paren{V_c^{(\n+1)}}^2\over2\paren{\n!}^2{V_c}^2}\,{\delta\varphi}^{2\n}
		+O\fn{{\delta\varphi}^{2\n+1}},	\label{epsilon}\\
\eta
	&:=	M_P^2{V''\over V}
	=	{M_P^2\,V_c^{(\n+1)}\over\paren{\n-1}!\,V_c}\,{\delta\varphi}^{\n-1}
		+O\fn{{\delta\varphi}^{\n}},	\label{eta}\\
\zeta^2
	&:=	M_P^4{V'''V'\over V^2}
	=	{M_P^4\paren{V_c^{(\n+1)}}^2\over\paren{\n-2}!\,\n!\,V_c^2}\,
			{\delta\varphi}^{2\n-2}
		+O\fn{{\delta\varphi}^{2\n-1}}. \label{zeta}
}
We see that $\epsilon\ll\ab{\eta},\zeta^2$ for $\delta\varphi\ll M_P$. 
The inflation ends when $\epsilon$ becomes of order unity, and we define its end point by $\epsilon(\delta\varphi_\text{end})=1$ to get
\al{
\paren{\delta\varphi_\text{end}}^\n
	&\simeq	{\sqrt{2}\,\n!\,V_c\over M_PV_c^{(\n+1)}}.
}
The e-folding number $N$ from a given stage of the inflation $\varphi=\varphi_c+\delta\varphi$ to its end $\varphi_\text{end}=\varphi_c+\delta\varphi_\text{end}$ is
\footnote{
The problem about the initial condition can be avoided by considering the eternal inflation scenario at the saddle point~\cite{Hamada:2015ria}.
}
\al{
N	=	\int_{\varphi_\text{end}}^\varphi{\df\varphi\over M_P^2}{V\over V'}
	&=	{\n!\over\paren{\n-1}}{V_c\over M_P^2V^{(\n+1)}_c}\sqbr{{1\over\paren{\delta\varphi_\text{end}}^{\n-1}}-{1\over\paren{\delta\varphi}^{\n-1}}}\nn
	&\simeq
		{\n!\over\paren{\n-1}}{V_c\over M_P^2\ab{V_c^{(\n+1)}}\ab{\delta\varphi}^{\n-1}},
			\label{e-folding}
}
where we have assumed $\ab{\delta\varphi_\text{end}}\gg\ab{\delta\varphi}$ in the last step.
From Eqs.~\eqref{epsilon}\eqref{eta}\eqref{zeta}\eqref{e-folding}, we obtain
\al{
\epsilon
	&=	{1\over2M_P^2}\sqbr{
		{\n!\over\paren{\n-1}^{\n}}
		{1\over N^{\n}}
		{V_c\over M_P^2\ab{V_c^{(\n+1)}}}
		}^{{2\over\n-1}},	&
\eta
	&=	-{\n\over\paren{\n-1}N},	&
\zeta^2
	&=	{\n\over\paren{\n-1}N^2}.
		\label{gen slow roll}
}

The cosmological observables, namely the scalar perturbation $A_s$, spectral index $n_s$, tensor-to-scalar ratio $r$, and running index ${\df n_s/\df\ln k}$
\footnote{
It appears that these quantities change their values discretely with $\n$. This is because $\n$ is the number of fine-tunings. However, if we take the next order term into account, we can explicitly check that the limit of $V_c^{(\n+1)}\to0$ continuously connects the case $\n$ to $\n+1$. Thus we can have fractional $\n$ effectively.
}
\al{
A_s
	&=	{V\over24\pi^2\epsilon M_P^4}
		\label{scalar perturbation}, \\
n_s
	&=	1-6\epsilon+2\eta
		\label{spectral index}
	\simeq
		1-{2\n\over\paren{\n-1}N},	\\
r	&=	16\epsilon,	\\
{\df n_s\over\df\ln k}
	&=	-16\epsilon\eta+24\epsilon^2+2\zeta^2
	\simeq2
		{\n\over\paren{\n-1}N^2},
		\label{running index}
}
are constrained by the Plnack 2015 data~\cite{Ade:2015lrj}
\al{&
A_s\simeq2.2\times 10^{-9},&
0.954<n_s&<0.980,&
r	&<	0.168,&
-0.03<{\df n_s\over\df\ln k}
	&<	0.007,
\label{Planck data}
}
at the 95\% CL.\footnote{
To give the most conservative bound, here we employ the constraint from the Planck TT+lowP  data.
}
The e-folding number
\al{
N_*
	&=	62-\ln\fn{10^{16}\GeV\over V_\text{end}^{1/4}}
	\simeq
		64+{1\over4}\ln\epsilon
		\label{COBE e-folding}
}
corresponds to the stage of inflation observed by the Planck experiment.
%
We note that this model gives a concave potential , $\eta<0$, which is favored by the recent Planck data.\footnote{
We thank the referee for pointing out this point.
}
%



\section{Saddle point inflation in string-inspired theory}\label{Sec:string}
In this section, we consider the saddle point inflation in the non-supersymmetric heterotic-like string model. 
Here we assume that the tree level potential of the inflaton is absent.
This is realized if the inflaton comes from the extra component of the gauge field/metric,  for example.
Then the dominant contribution to the potential is the one loop correction, 
which is suppressed compared to the string scale by the loop factor:
\al{
\int {d^dk\over(2\pi)^d}
={S_{d-1}\over2(2\pi)^d}\int dk^2 (k^2)^{{d\over2}-1}
\sim {S_{d-1}\over2(2\pi)^d} M_s^{d}
.
}
For $d=10$, we obtain the following numerical value
\al{
C_{\text{loop}}\equiv{S_{d-1}\over2(2\pi)^d}={2\pi^5\over\Gamma(5)}{1\over2(2\pi)^{10}}
\simeq 1.3 \times 10^{-7}.
}
In fact, the 10 dimensional cosmological constant of $SO(16)\times SO(16)$ heterotic string theory~\cite{SO(16)} is calculated as
\al{
\Lambda_{SO(16)\times SO(16)}\simeq 3.9\times 10^{-6} M_s^{10}.
}

Because we assume that the tree potential of the inflaton vanishes, the effective action below the string scale becomes
\al{
S&=
{M_s^8\over g_s^2}\int d^{10}x\sqrt{g} A(\chi) \mathcal{R}
+{M_s^8\over g_s^2}\int d^{10}x\sqrt{g} B(\chi) (\partial \chi)^2 
+C_{\text{loop}} M_s^{10}\int d^{10}x\sqrt{g} V(\chi)+\cdots\nn
&
=
{M_s^8\over g_s^2} V_6 \int d^{4}x\sqrt{g} A(\chi) \mathcal{R}
+{M_s^8\over g_s^2} V_6 \int d^{4}x\sqrt{g} B(\chi) (\partial \chi)^2 
+C_{\text{loop}} M_s^{10} V_6 \int d^{4}\sqrt{g} V(\chi)+\cdots.
}
Here $\chi$ is the dimensionless inflaton field, $g_s$ is the string coupling, $V(\chi)$ is the one loop potential, and $V_6$ is the compactification volume.
Because $M_s$ is the only mass scale of the theory,
$A(\chi)$, $B(\chi)$ and $V(\chi)$ should be functions of order one
\al{
&
A(\chi)=a_0+a_2 g_s^2 \chi^2+\cdots,
&
B(\chi)=b_0+b_2 g_s^2 \chi^2+\cdots,
&
\,\,
V(\chi)=v_0+v_2 g_s^2 \chi^2+\cdots,
&
}
with $a_i$'s, $b_i$'s and $v_i$'s being order one constants.
Next let us move to the Einstein frame. 
Namely, we redefine the metric in such a way that $A(\chi)$ becomes 1. 
In the Einstein frame, we have
\al{
S&=
M_P^2 \int d^{4}x\sqrt{g} \mathcal{R}
+M_P^2 \int d^{4}x\sqrt{g} C(\chi) (\partial \chi)^2
+C_{\text{loop}} g_s^2 M_P^2 M_s^2 \int d^{4}x\sqrt{g} U(\chi).
}
Here
\al{&
M_P^2={M_s^2\over g_s^2}(M_s^6 V_6),
&
C(\chi)=c_0+c_2 g_s^2 \chi^2+\cdots,
&
\,\,
U=u_0+u_2 g_s^2 \chi^2+\cdots,
&
}
where $c_i$'s and $u_i$'s are order one constants.
In terms of the dimensionless canonical field $\phi$, the action becomes
\al{
S=M_P^2\int d^4x \sqrt{g} \, \mathcal{R}
+M_P^2\int d^4x \sqrt{g} \, (\partial \phi)^2
+C_{\text{loop}} {g_s^2 M_P^2 M_s^2}\int d^4x \sqrt{g} \, W\left({\phi}\right),
}
where $W(\phi)$ is a function of order one.

The argument so far is quite general.
In the following, we assume that the potential has a saddle point where the first $\n$ derivatives vanish as in Sec.~\ref{Sec:saddle point inflation}.
This may happen by some mechanism beyond the ordinary local field theory such as the multiple point criticality principle~\cite{MPP} and the maximum entropy principle~\cite{MEP}.
%
%
Here, we take
\al{
W(\phi)&=W_0\left(1-\left(1-{\phi\over\phi_c}\right)^{\n+1}\right),
}
as a simple possibility. 
%
We expect that $\phi_c$ is the order one quantity.
In terms of the canonical field $\varphi=M_P \phi$, the potential $V(\varphi)$ becomes
\al{\label{saddle point potential}
V(\varphi)&=C_{\text{loop}} {g_s^2 M_P^2 M_s^2}\times W\left({\varphi\over M_P}\right),
}
%
Then, from Eq.~\eqref{gen slow roll}, we get
\al{
\epsilon
	&=	{1\over2M_P^2}\sqbr{
		{\n!\over\paren{\n-1}^{\n}}
		{1\over N^{\n}}
		{V_c\over M_P^2\ab{V_c^{(\n+1)}}}
		}^{{2\over\n-1}}\nn
	&=	{1\over2}\sqbr{
		{\n!\over\paren{\n-1}^{\n}}
		{1\over N^{\n}}
		{\phi_c^{\n+1} \over(\n+1)!}
		}^{{2\over\n-1}}.\label{eq:epsilon}
}
%
Furthermore, Eq.~\eqref{eq:epsilon} and the COBE normalization Eq.~\eqref{scalar perturbation} fix the value of $V_c=C_\text{loop}g_s^2M_P^2M_s^2 W_0$, from which we  can obtain the string scale.
In Table~$1$
, we present the predictions of the cosmological parameters taking $C_\text{loop}=10^{-7}, N=60$. 
\begin{table}\label{Table:predictions}
\begin{center}
\begin{tabular}{|c|c|c|c|c|}
\hline
$\n$&$n_s$&$\epsilon$&$V_c/M_P^4$&$g_s M_s\sqrt{W_0}/M_P$\\ \hline
$2$&$0.933\cdots$&$4.3\times10^{-9}\phi_c^6$&$2.2\times10^{-15}\phi_c^6$&$1.5\times10^{-4}$\\ \hline
$3$&$0.95$&$7.2\times10^{-8}\phi_c^4$&$3.8\times 10^{-14}\phi_c^4$&$6.1\times10^{-4}$\\ \hline
$4$&$0.955\cdots$&$1.7\times10^{-7}\phi_c^{10/3}$&$8.6\times10^{-14}\phi_c^{10/3}$&$9.3\times10^{-4}$\\ \hline
$5$&$0.95833\cdots$&$2.3\times10^{-7}\phi_c^3$&$1.2\times 10^{-13}\phi_c^3$&$1.1\times10^{-3}$\\ \hline
$6$&$0.96$&$2.6\times10^{-7}\phi_c^{14/5}$&$1.4\times 10^{-13}\phi_c^{14/5}$&$1.2\times10^{-3}$\\ \hline
\end{tabular}
\caption{The predictions of cosmological parameters and the string scale for $N=60, C_\text{loop}=10^{-7}$.}
\end{center}
\end{table}
From this table, we can see that $\n\geq4$ is favored by the current observation Eq.~\eqref{Planck data}. 
The tensor to scalar ratio is very small compared to the current limit provided that $\phi_c$ is of order one.
As we vary $\n$ from $2$ to $6$, $g_s M_s$ takes from $4\times 10^{14}$ GeV to $3\times 10^{15}$ GeV for $W_0=1$.
If $g_s$ takes $\mathcal{O}(0.1)$, the result indicates that the $M_s$ is around the GUT scale~$\sim10^{16}$ GeV.

\begin{figure}
\begin{center}
\includegraphics[width=10cm]{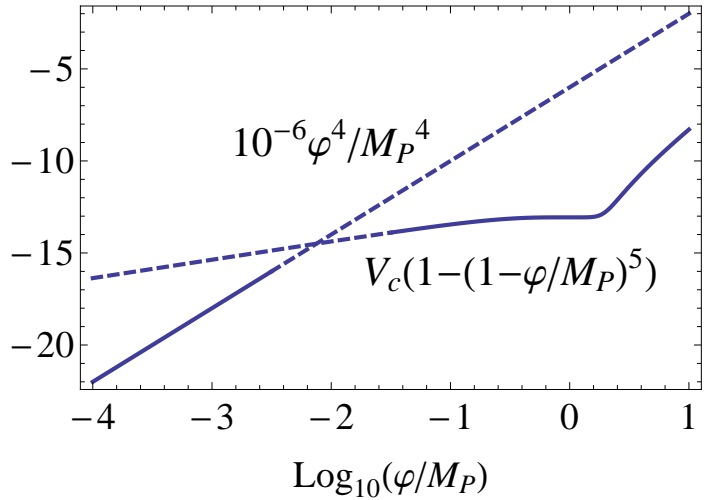}
\end{center}
\caption{Matching between Eq.~\eqref{saddle point potential} and Eq.~\eqref{Higgs potential}.}
\label{fig:matching}
\end{figure}

\begin{figure}
\begin{center}
\includegraphics[width=10cm]{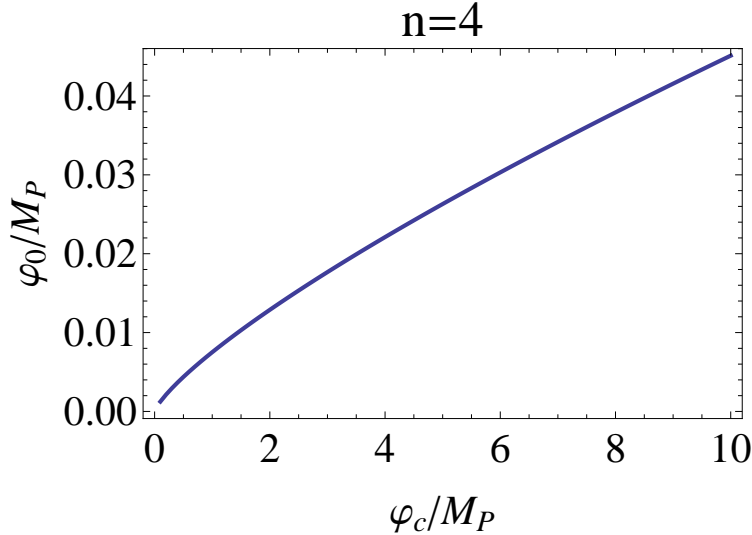}
\end{center}
\caption{$\varphi_0$ as a function of $\varphi_c$. $\varphi_0$ is the value of $\varphi$ for which $V_\text{SM}$ equals $V$.}
\label{fig:chi0}
\end{figure}

%
Finally let us discuss the possibility of identifying the inflaton as the SM Higgs.
The recent analysis shows that the Higgs potential for the large values of the Higgs field $h$ is roughly given by
\al{\label{Higgs potential}
V_\text{SM}\sim 10^{-6}h^4,
}
in the SM~\cite{Bare mass,SM stability} and its simple extensions~\cite{extension} when the top mass is around $171\text{--}172\GeV$.
We examine whether $V_\text{SM}$ can be connected to the potential $V$ in Eq.~\eqref{saddle point potential} under the assumption that $\varphi$ is identified as $h$.
In Fig.~\ref{fig:matching}, Eqs.~\eqref{saddle point potential}\eqref{Higgs potential} are plotted.
Here we take $\n=4, \phi_c=1, W_0=1$ as an example.
One can see that two lines are crossed at around $\varphi\simeq10^{16}\GeV$, which we call $\varphi_0$.
We interpret this as the indication that the potential is given by the SM at lower energies, and becomes stringy, Eq.~\eqref{saddle point potential}, above the string scale~$\sim10^{16}\GeV$.
We also show $\varphi_0$ as a function of $\varphi_c=\phi_c M_P$ in Fig.~\ref{fig:chi0}.
$\varphi_0$ takes the order of $10^{16}\GeV$ for $\varphi_c=\mathcal{O}(M_P)$.

%
%
%
%


\section{Summary}\label{Sec:summary}
We have examined the possibility of the saddle point inflation in the context of non-supersymmetric string theory, which is ubiquitous and becomes more realistic in light of the recent LHC result.
Contrary to supersymmetric theory, the potential is generated perturbatively.
We have assumed that the potential of the inflaton is identically zero at the tree level, and it is radiatively generated by the loop effect.
We have estimated the string scale that realizes a successful inflation assuming that the potential is tuned so that it has a saddle point where first $\n$ derivatives vanish.
Interestingly, the string scale becomes around the GUT scale, $\sim10^{16}$GeV, if the string coupling is $\mathcal{O}(0.1)$.
Furthermore, we have found that it is reasonable to identify the inflaton as the Higgs field.
%
%
It is interesting that, 
in addition to the LHC results, 
the scale of the inflation supports non-supersymmetric string theory.

\subsection*{Acknowledgement}
We thank Kin-ya Oda for useful discussions.
This work is supported by the Grant-in-Aid for Japan Society for the Promotion of Science (JSPS) Fellows No.25·1107 (YH) and No.27·1771 (KK).
H.K.’s work is supported in part by the Grant-in-Aid for Scientific Research No. 22540277.

\end{document}